\documentstyle[epsf]{mn}
\newcommand{\bm}[1]{{\bf#1}}

\title{Numerical study of statistical properties of the lensing
excursion angles}
\author[T. Hamana \& Y. Mellier]
{Takashi Hamana${}^{1,2}$ and Yannick Mellier${}^{1,3}$\\
$^1$Institut d'Astrophysique de Paris, CNRS, 98bis Boulevard Arago, F75014
PARIS, France\\
$^2$Max-Planck-Institut f\"ur Astrophysik,
P.O.~box 1317, D--85741 Garching, Germany\\
${}^3$Observatoire de Paris, DEMIRM, 61 avenue de l'Observatoire,
75014 PARIS, France}
\date{Accepted ....; Received ....; in original form ....}
\pagerange{\pageref{firstpage}--\pageref{lastpage}}
\pubyear{....}
\begin{document}
\maketitle
\label{firstpage}

\begin{abstract}
We present ray tracing simulations combined with sets of large N-body
simulations.
Experiments were performed to explore, for the first time,
statistical properties of fluctuations in angular separations of
nearby light ray pairs (the so-called lensing excursion angle) induced by
weak lensing by large-scale structures.
We found that the probability distribution function (PDF) of the
lensing excursion angles is not simply Gaussian but has an exponential
tail. 
It is, however, found that the tail, or more generally non-Gaussian
nature in the PDF has no significant impact on the weak lensing of
the CMB. 
Moreover, we found that the variance in the lensing excursion angles predicted
by the power spectrum approach is in good agreement with our numerical
results.
These results demonstrate a validity of using the power spectrum
approach to compute lensing effects on the CMB.
\end{abstract}

\section{Introduction}
Weak lensing effects on the temperature anisotropy in the
cosmic microwave background (CMB) have been recognized as a powerful
probe of cosmology (see Mellier 1999; Bartelmann \& Schneider 2001
for reviews).
Dark matter distribution along the line of sight between the last 
scattering surface (LSS) and us deflects the light ray trajectories 
and induces distortions in the pattern of the CMB anisotropies.
Since the gravitational lensing is directly sensitive to the matter
distribution up to the LSS, lensing signatures imprinted on the CMB may
provide important information about the matter distribution on large
scales and at high redshifts.
In this point of view, various methods have been proposed 
(Seljak 1996; Bernardeau 1997; 1998; Metcalf \& Silk 1997; Zaldarriaga
\& Seljak 1998; 1999; Zaldarriaga 2000; Suginohara, Suginohara \&
Spergel 1998; Van Waerbeke, Bernardeau \& Benabed 2000; Takada,
Komatsu \& Futamase 2000; Takada \& Futamase 2001).
These lensing signatures are generally small but are measurable with two
planned satellite missions, MAP\footnote{See the MAP homepage at
http://map.gsfc.nasa.gov.} and Planck\footnote{See the Planck homepage
at http://astro.estec.esa.nl/Planck.}, and can help to break some of
the parameter degeneracies in the CMB (Bernardeau 1997; Metcalf \&
Silk 1998; Takada \& Futamase 2000).
The change in a separation angle of two nearby light rays caused by
weak lensing (lensing excursion angle) plays a key role in
studying weak lensing of the CMB, especially weak lensing effects
on two-point statistics of the CMB.

The analytical prediction of the weak lensing effect on the CMB power
spectrum in modern cosmological models was first developed by 
Seljak (1994) based on linear perturbation theory (the so-called
power spectrum approach).
While, in his subsequent paper, Seljak (1996) examined effects of the
nonlinearity in the density
through analytic fitting formulae (Peacock \& Dodds, 1996), and
pointed out that the nonlinearity is very important on sub-degree
scales.
Furthermore statistics of the lensing excursion angles due to weak
lensing are frequently assumed to be Gaussian without a
rigorous basis (e.g., Seljak 1996).
Numerical simulations are, therefore, needed for testing the validity
and limitation of the semi-analytic approach.

The purpose of this paper is to examine the statistical properties of
the lensing excursion angles using ray-tracing simulations combined
with large $N$-body simulations for the first time.
Since the simulations were originally constructed for the cosmic shear
statistics (Van Waerbeke et al.~2001; Hamana et
al.~2001b) the maximum redshift is taken by $z\sim3$, which is about a half
way to the LSS.
This choice may not seem to be enough for studying the weak lensing
effects on the CMB.
However, most of the contributions to the CMB lensing
come from structures at $z<3$.
To take an example shown by Suginohara et al.~(1998), in the COBE
normalized SCDM model, the structures within $z<3$ contribute 85\% to
the variance of the lensing angular excursion from the LSS (see Figure
1b of their paper).
The remaining 15\% will not significantly alter the basic properties
of weak lensing of the CMB, and therefore we can, at least, study its
essential features.
It should be, however, noticed that this 15\% is surely the most
interesting for probing the early stage of the large-scale structure
formation with weak lensing of the CMB.

The outline of this paper is as follows.
Our models and method of the ray-tracing simulation are
described in \S2.
In \S3, we show the numerical resolution of our simulation and discuss
its limitation.
Using results of the ray-tracing simulations, we examine statistical
properties of the lensing excursion angles in \S4.
We conclude in \S5.

Throughout this paper, we work in the comoving coordinates system.
The cosmological parameters are denoted with usual notations; Hubble
constant, $H_0=100h$km/s/Mpc; the density parameter, $\Omega_{\rm m}$;
the cosmological constant, $\Omega_\Lambda$; the variance of the
density fluctuation in a sphere of radius $8h^{-1}$Mpc, $\sigma_8^2$.

\section{Numerical methods and models}
\label{sec:numerical}

In this section, we describe methods of the $N$-body simulation
(\S \ref{sec:pm}) and the ray-tracing simulation (\S \ref{sec:rt}),
and summarize our models.

\subsection{$N$-body simulation and the tiling technique}
\label{sec:pm}

\begin{table}
\caption{Cosmological parameters.}
\label{table:cosmo}
\begin{center}
\begin{tabular}{ccccc}
\hline
Model & $\Omega_{\rm m}$ & $\Omega_\lambda$ & $h$ & $\sigma_8$ \\
\hline
SCDM & 1.0 & 0.0 & 0.5 & 0.6 \\
OCDM & 0.3 & 0.0 & 0.7 & 0.85 \\
$\Lambda$CDM & 0.3 & 0.7 & 0.7 & 0.9 \\
\hline
\end{tabular}
\end{center}
\end{table}

The $N$-body simulation data set were generated with a PM code (see
Hamana et al.~2001b for a detail description). 
Each $N$-body experiment involves $256^2\times512$ particles in a
periodic rectangular box of size $(L,L,2L)$.
The mesh used to compute the forces was $256^2\times 512$. 
The initial conditions are generated adopting the transfer function of
Bond \& Efstathiou (1984) with the shape parameter $\Gamma=\Omega_{\rm m} h$.
We adopt three cosmological models; two flat models with and without
the cosmological constant and one open model.
The amplitude of the power spectrum is normalized by the cluster
abundance (Eke, Cole \& Frenk 1996; Kitayama \& Suto 1997).
In Table \ref{table:cosmo}, cosmological parameters in each model are
summarized.

\begin{figure}
\begin{center}
\begin{minipage}{8cm}
\begin{center} 
\epsfxsize=8cm \epsffile{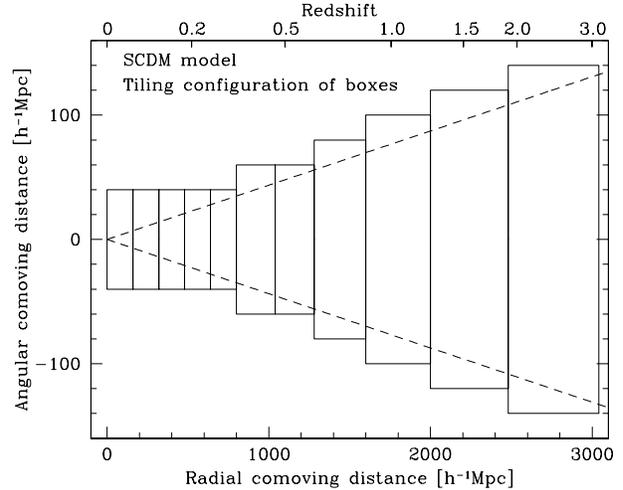}
\end{center}
\end{minipage}
\caption{Tiling configuration of $N$-body simulation boxes for SCDM
model. Dashed lines (cone) show the comoving angular diameter
distance of $\pm2.5$ arcmin.
Note that the scale of the vertical axis is exceedingly enlarged for
clarity, the long side of the rectangular boxes is parallel to the
horizontal axis.}
\label{fig:tiling}
\end{center}
\end{figure}

A light-cone of the particles was extracted from each simulation
during the run as explained in Hamana, Colombi \& Suto (2001a). 
Our aim was that the light-cone covers a large redshift range, 
$0 \leq z \la 3$, and a field of view of $5\times5$  square degrees
with a wide angular dynamic range.
To do that, we adopted the {\it tiling} technique first proposed by 
White \& Hu (2000).
We performed $N_{\rm box}=11$, 12 and 13 independent simulations for
SCDM,  OCDM and $\Lambda$CDM, respectively, covering adjacent redshift
intervals  $[z_i^{\rm min}, z_i^{\rm max}]$, $i=1,\ldots,N_{\rm box}$. 
The size of each simulation is chosen such that the portion of the
light-cone in $[z_i^{\rm min}, z_i^{\rm max}]$ 
(aligned with the third axis) fits the box-size as shown in
Figure \ref{fig:tiling}.
This way, angular resolution is approximately conserved as a function
of redshift, except close to the observer (to be discussed in \S
\ref{sec:resolution}).  
Finally, in order to keep large scale modes that may contribute to the
weak lensing, we impose the supplementary constraint $L\geq 80h^{-1}$Mpc. 
As a result, $L$ follows the sequence with redshift summarized in
Table \ref{table:boxes}.

\begin{table}
\caption{Caracteristics of PM simulation boxes.}
\label{table:boxes}
\begin{center}
\begin{tabular}{lccccc}
\hline
Model & $L${}$^{\rm{a}}$ & $m_{\rm part}${}$^{\rm{b}}$ &
\multicolumn{2}{c}{Redshift range} \\  
 & $h^{-1}$Mpc & $h^{-1}10^{10}M_\odot$ & $z_{\rm min}$ &
$z_{\rm max}$ \\
\hline
SCDM & 80 & 0.85 & 0.0 &  0.055545\\
 & 80 & 0.85 & 0.055545 & 0.11585 \\
 & 80 & 0.85 & 0.11585 & 0.18147 \\
 & 80 & 0.85 & 0.18147 & 0.25306 \\
 & 80 & 0.85 & 0.25306 & 0.33136 \\
 & 120 & 2.9 & 0.33136 & 0.46332 \\
 & 120 & 2.9 & 0.46332 & 0.61591 \\
 & 160 & 6.8 & 0.61591 & 0.85950 \\
 & 200 & 13 & 0.85950 & 1.2500 \\
 & 240 & 23 & 1.2500 & 1.9054 \\
 & 280 & 36 & 1.9054 & 3.1088 \\
 & & & &  \\
OCDM & 80 & 0.25 & 0.0 & 0.055010 \\
 & 80 & 0.25 & 0.055010 & 0.11355 \\
 & 80 & 0.25 & 0.11355 & 0.17589 \\
 & 80 & 0.25 & 0.17589 & 0.24235 \\
 & 80 & 0.25 & 0.24235 & 0.31326 \\
 & 120 & 0.86 & 0.31326 & 0.42880 \\
 & 120 & 0.86 & 0.42880 & 0.55664 \\
 & 160 & 2.0 & 0.55664 & 0.74917 \\
 & 200 & 4.0 & 0.74917 & 1.0323 \\
 & 240 & 6.9 & 1.0323  & 1.4508 \\
 & 320 & 16 & 1.4508 & 2.1935 \\
 & 440 & 42 & 2.1935 & 3.7702 \\
 & & & & \\
$\Lambda$CDM & 80 & 0.25 & 0.0 & 0.053997 \\
 & 80 & 0.25 & 0.053997 & 0.10942 \\
 & 80 & 0.25 & 0.10942 & 0.16641 \\
 & 80 & 0.25 & 0.16641 & 0.22515 \\
 & 80 & 0.25 & 0.22515 & 0.28581 \\
 & 120 & 0.86 & 0.28581 & 0.38084 \\
 & 120 & 0.86 & 0.38084 & 0.48137 \\
 & 160 & 2.0 & 0.48137 & 0.62541 \\
 & 200 & 4.0 & 0.62541 & 0.82485 \\
 & 240 & 6.9 & 0.82485 & 1.1001 \\
 & 280 & 11 & 1.1001 & 1.4870 \\
 & 360 & 23 & 1.4870 & 2.1341 \\
 & 440 & 42 & 2.1341 & 3.2942 \\
\hline
\end{tabular}
\end{center}
\begin{list}{}{}
\item[$^{\rm{a}}$] The short side length of boxes in comoving
units, the long side length is $2L$ in all cases.
\item[$^{\rm{b}}$] Particle mass.
\end{list}
\end{table}

\subsection{Ray-tracing simulation}
\label{sec:rt}

Let us first summarize basic equations of the multiple lens-plane
algorithm which are directly relevant to this paper (see
e.g.~Schneider, Ehlers \& Falco 1992 and Jain, Seljak \& White 2000 for
details).
We shall denote the comoving Cartesian coordinate system as
$(x_1,x_2,y)$ with $y$ being the third axis and the origin being at
the observer point.
Since the field of view we consider is small ($5\times 5$ square degrees),
the radial comoving distance (we shall denote it by $\chi$) can be
approximated to $y$. 
In the standard multiple-lens plane algorithm, the distance between
source and observer is divided into $N$ intervals separated by
comoving distance $\Delta y$.
The matter content in each interval is projected onto {\it lens
planes} perpendicular to $y$-axis.
The projected density contrast of $i$-th plane is defined by,
\begin{eqnarray}
\label{surfacedens}
\delta^{\rm proj}_i(x_1,x_2) = \int_{y_{i-1}}^{y_i} dy\, \delta(x_1,x_2,y), 
\end{eqnarray}
where $\delta=\rho/\bar{\rho}-1$, $y_i$ denotes $y$-position
of $i$-th lens plane.
The two-dimensional deflection potential of $i$-th lens plane is
related to the projected density contrast via the
two-dimensional Poisson equation by
\begin{eqnarray}
\label{2dimpot}
\nabla^2\Psi^i(x_1,x_2) & =&{{8 \pi G \bar{\rho}}\over{3c^2}}
\delta_i^{\rm proj}(x_1,x_2)\nonumber\\
&=& 3 \Omega_{\rm m} \left({{H_0}\over c}\right)^2
\delta_i^{\rm proj}(x_1,x_2)
\end{eqnarray}
The light ray position on $n$-th lens plane of a ray with image
positions $\bmath{\theta}^1$ are computed using the
multiple lens-plane equation,
\begin{eqnarray}
\label{lens-eq}
\bmath{\theta}^n = \bmath{\theta}^1-\sum_{i=1}^{n-1}
{{f(\chi_n-\chi_i)}\over{a(\chi_i) f(\chi_n)}} \bm{\nabla}_\perp
\Psi^i,
\end{eqnarray}
where $f(\chi)$ denotes the comoving angular diameter distance,
defined as $f(\chi)=K^{-1/2} \sin K^{1/2} \chi$, $\chi$, $(-K)^{-1/2}
\sinh (-K)^{1/2} \chi$ for $K>0$, $K=0$, $K<0$, respectively, where
$K$ is the curvature which can be expressed as
$K=(H_0/c)^2(\Omega_{\rm m}+\Omega_\lambda-1)$,
and $\bmath{\nabla}_\perp$ denotes either $\partial/\partial x_1$ or
$\partial/\partial x_2$.
Note that the spatial position $\bmath{x}^i$ on $i$-th plane is related
to the angular position by $\bmath{x}^i=f(\chi_i)\bmath{\theta}^i$. 
The evolution equation of the
Jacobian matrix, which describes deformation of an infinitesimal light ray
bundle, is written by
\begin{eqnarray}
\label{jacobi}
\bmath{A}_n = \bmath{I}-\sum_{i=1}^{n-1}
{{f(\chi_i) f(\chi_n-\chi_i)}\over{a(\chi_i) f(\chi_n)}} 
\bmath{U}_i \bmath{A}_i,
\end{eqnarray}
where $\bmath{I}$ denotes the identity matrix, and
$\bmath{U}_i$ is the, so-called, optical tidal matrix defined by
\begin{eqnarray}
\label{U}
\bmath{U}_i =
\left(
\begin{array}{cc}
\Psi_{,11}^i & \Psi_{,12}^i\\
\Psi_{,21}^i & \Psi_{,22}^i
\end{array}
\right),
\end{eqnarray}
with commas denoting the differentiation with respect to $\bmath{x}$.
It should be noticed that the first and second derivatives of
$\Psi$ in equations (\ref{lens-eq}) and (\ref{jacobi}) must
be evaluated at ray positions computed by the lens equation
(\ref{lens-eq}). 
The Jacobian matrix is usually decomposed into,
\begin{eqnarray}
\label{A}
\bmath{A} =
\left(
\begin{array}{cc}
1-\kappa-\gamma_1 & -\gamma_2-\omega\\
-\gamma_2+\omega & 1-\kappa+\gamma_1
\end{array}
\right),
\end{eqnarray}
where $\kappa$ represents convergence,
$|\gamma|=(\gamma_1^2+\gamma_2^2)^{1/2}$ is the amplitude of shear of
light ray bundle, and $\omega$ is a net rotation of the beam.

We adopt a fixed interval between lens planes by
$\Delta y=80h^{-1}$Mpc (for this choice, the long side length of simulation
boxes are chosen so that they are multiples of $80h^{-1}$Mpc).
The procedure to trace light
rays through $N$-body data with the multiple lens-plane algorithm can
be described as follows: 
(i) computing the projected density contrast, $\delta_i^{\rm proj}$
from particle distribution in $N$-body data. (ii) computing the
two-dimensional deflection potential, $\Psi^i$ via Poisson equation
(\ref{2dimpot}) and evaluating the first and second derivatives of it.
(iii) evaluating the distance combination for each lens plane and
performing the summation in equations (\ref{lens-eq}) and (\ref{jacobi}).
In the rest of this section, we describe each step in some detail.

(i) A simulation particle located at $y$-coordinate between
$y_{i-1}$ and $y_i$ is projected onto the $i$-th lens plane at $y_i$.
This projection is done in parallel to the third axis, and thus the
particle's $(x_1,x_2)$ positions are not changed.
Since the $N$-body data is periodic in $\bmath{x}$
directions\footnote{$N$-body simulation data is not periodic in $y$
direction, because
we used the light-cone output.}, the projected particle distributions
are also periodic. 
The projected density field is computed on a $512^2$ square lattice
from the projected particle distribution, using the triangular shaped
cloud (TSC) assignment scheme (Hockney \& Eastwood, 1988).
In this step, the most important point is the choice of both the
smoothing scheme and the grid size. 
Our main consideration when choosing them was to
maintain the resolution provided by the $N$-body simulation and at the
same time to remove shot noise due to discreteness in $N$-body
simulations by a relevant smoothing.
We tested the smoothing scale by varying the size of grids, $256^2$,
$512^2$ and $1024^2$.
In the case of $1024^2$ lattice, we found a white-noise contribution to
the power spectrum of the lensing convergence at small scales, which
is a typical signature
of the discreteness effect (Jain et al.~2000).
However, the discreteness effect was smoothed sufficiently well in both
$256^2$ and $512^2$ lattice cases, so the white-noise
contribution was not detected for those cases.
On the other hand, we also found that the small scale power was damped
significantly for the $256^2$ case, which indicates too much smoothing.
We also tested another smoothing scheme, cloud-in-cell (Hockney \&
Eastwood, 1988), with the same three grid sizes. 
In these cases, we found a non-negligible white-noise contribution. 
We, therefore, decided to adopt TSC with $512^2$ lattice.

(ii) Poisson equation (\ref{2dimpot}) is solved to compute $\Psi^i$
using the fast Fourier transform method with the periodic boundary
condition.
The first and second derivatives of $\Psi^i$ are evaluated on the
lattice points using the usual finite difference method (see Appendix
A of Premadi, Martel \& Matzner 1998 for explicit expressions).

(iii) $512^2$ rays are traced backward from the observer point.
The initial ray directions are set on $512^2$ grids, which correspond
to pixels of angular size $5\degr/512\sim 0.59$ arcmin.
For each ray, we first computed ray positions on all lens planes in an
iterative manner, using the lens equation (\ref{lens-eq}).
The first and second derivatives of $\Psi^i$ on a ray position are
linearly interpolated from four nearest grids on which they were
pre-computed (step (ii)).
Finally, the summation in equation (\ref{jacobi}) is performed.
The light ray positions and four components of the Jacobian matrix on
desired source planes are stored.
Exact values of the source redshifts used in this paper are summarized
in Table \ref{table:zs}.

\begin{table}
\caption{Exact values of the source redshifts. In the text, they
are refereed to as $z_s=1$, 2 and 3.}
\label{table:zs}
\begin{center}
\begin{tabular}{lccc}
\hline
{} & SCDM & OCDM & $\Lambda$CDM \\
\hline
$z_1$ & 1.0025 & 1.0323 & 1.0034\\
$z_2$ & 2.0422 & 1.9836 & 1.9721\\
$z_3$ & 3.1088 & 2.9501 & 3.0419\\
\hline
\end{tabular}
\end{center}
\end{table}

We performed 40 realizations of the underlying density field for each
cosmological model by random shifts of the simulation boxes in the
$(x_1,x_2)$ directions using periodic boundary condition.
Note that lens planes coming from the same box are shifted in the same
way to maintain the clustering of matter in the box. 
It is important to note that the 40 realizations are not rigorously
independent because they come from the same set of the simulation boxes.
It is expected that the cosmic variance can not be correctly included
because of this lack of independence, 
but this does not change our results because on angular scales of our
interest (below 100 arcmin) it is very small.
 
\section{Numerical resolution}
\label{sec:resolution}

\begin{figure}
\begin{center}
\begin{minipage}{8cm}
\begin{center} 
\epsfxsize=8cm \epsffile{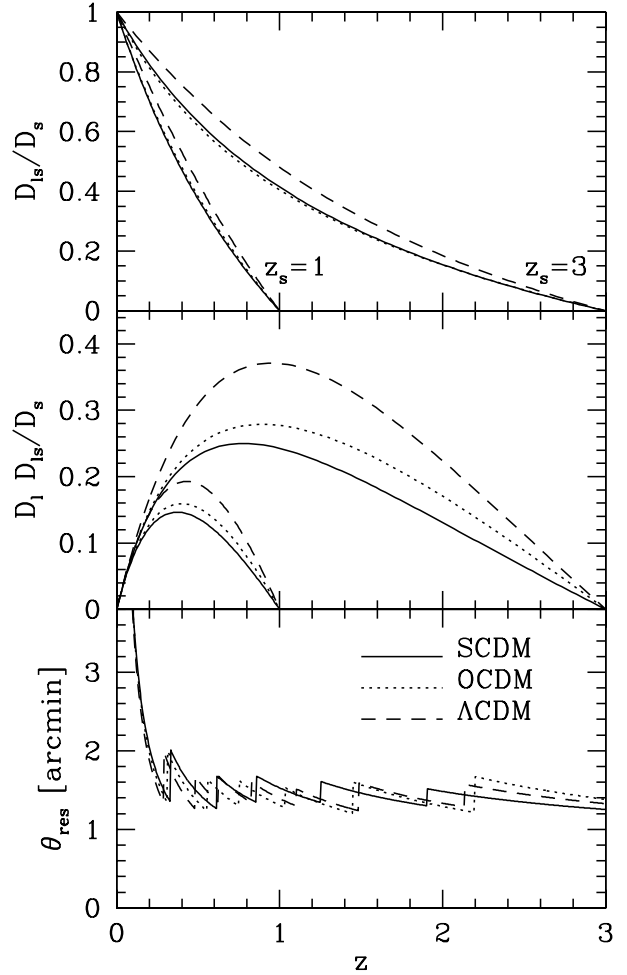}
\end{center}
\end{minipage}
\caption{{\it Top panel}: Distance combination appearing in the
equation (\ref{lens-eq}) for $z_s=1$ and 3 as a function of lens
redshift. 
{\it Middle panel}: Distance combination appearing in the
equation (\ref{jacobi}) (normalized by Hubble length, $c/H_0$) for
$z_s=1$ and 3 as a function of lens redshift.
{\it Bottom panel}: $\theta_{\rm res}$ as a function of redshift (see
text).}
\label{fig:resolution}
\end{center}
\end{figure}

The angular resolution of the ray-tracing simulation is basically
limited by the spatial resolution of the $N$-body simulations.
Since we used PM $N$-body code, the spatial resolution is simply
$r_{\rm res}\simeq L/256$.
This can be converted into the angular scale using the angular
diameter distance relation, $r_{\rm res}=f(\chi) \theta_{\rm res}$.
The bottom panel of Figure \ref{fig:resolution} shows such angular
scales computed for three cosmological models.
Thanks to the tiling technique, $\theta_{\rm res}$ is almost constant
($\theta_{\rm res}\sim 1.5$) at redshifts higher than 0.2.
It increases as $\theta_{\rm res} \propto \chi^{-1}$ at $z\la 0.2$, 
because of the constraint on the minimum box size. 
The impact of this worse resolution on a measurement depends on the
quantity one wants to examine.
Roughly speaking, quantities which come from solving the lens equation 
(\ref{lens-eq}) are more influenced by the worse resolution than ones
related to the Jacobian matrix obtained via equation (\ref{jacobi}).

\begin{figure}
\begin{center}
\begin{minipage}{8.5cm}
\begin{center} 
\epsfxsize=8.5cm \epsffile{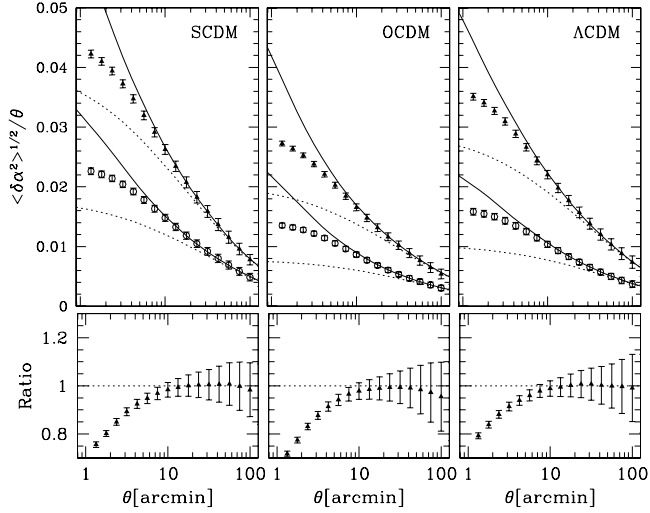}
\end{center}
\end{minipage}
\caption{{\it Top}: The root-mean-square of lensing excursion
angles between nearby light ray pairs divided by their intrinsic
separation $\theta$.
$\delta\alpha$ is either $\delta\alpha_1$ or $\delta\alpha_2$, and the
averaged value is plotted.
The open circles are for measurements of $z_s=1$, while the filled
triangles are for $z_s=3$.
The error bars denote the standard deviation computed among 40
realization.
The solid and dotted lines represent the nonlinear and linear
prediction, respectively.
{\it Bottom}: Ratio of the measurements to the nonlinear prediction
for $z_s=3$.
}
\label{fig:deltheta}
\end{center}
\end{figure}

The lensing excursion angle plays a crucial role on studies of the weak
lensing effects on the CMB and is defined by the difference in
deflection angles between nearby two light rays.
Each light ray trajectory is computed by the lens equation.
As the expression, eq.~(\ref{lens-eq}), indicated, the distance
combination $f(\chi_n-\chi_i)/f(\chi_n)$ acts as a weight function.
The top panel of Figure \ref{fig:resolution} shows this distance
combination as a function of the lens redshift for two cases, $z_s=1$
and 3, and clearly indicates that lensing deflections due to
structures at low redshifts are more weighted than those by high
redshift structures.
It can be, therefore, expected that the worse angular resolution at
low redshifts has a significant impact on the effective
accuracy of the light ray trajectory.
In order to check it, we compute the variance in the lensing excursion
angles of the two nearby light rays which would be observed with
a separation $\theta$ if there was no lensing. 
The deflection angle of a light ray is simply the difference between its
angular position in the first lens plane and that in the source
plane, i.e., $\bmath{\alpha}=\bmath{\theta}^1-\bmath{\theta}^n$.
The lensing excursion angle between nearby rays $A$ and $B$ is simply 
$\delta \bmath{\alpha}= \bmath{\alpha}^A - \bmath{\alpha}^B$,
and their intrinsic separation is $\theta=|\bmath{\theta}^{n,A} -
\bmath{\theta}^{n,B}|$.
In Figure \ref{fig:deltheta}, the results are plotted together with
the semi-analytic predictions by the power spectrum approach (Seljak
1994; 1996).
The measurements agrees very well with the nonlinear semi-analytic
prediction at scales larger than 10 arcmin.
Below that scale, the measurements are depressed reflecting the
effective resolution limit.
We may, therefore, conclude that the angular range, where our analysis
related to the lensing excursion angle is reliable, is between 10 arcmin
and 2 degree.

\begin{figure}
\begin{center}
\begin{minipage}{8.5cm}
\begin{center} 
\epsfxsize=8.5cm \epsffile{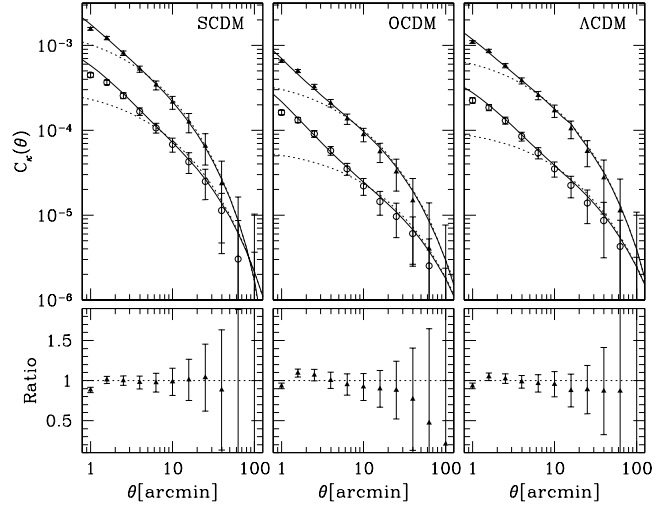}
\end{center}
\end{minipage}
\caption{Same as Figure \ref{fig:deltheta} but for two-point correlation
function of the lensing convergence.}
\label{fig:tc}
\end{center}
\end{figure}

Although we are not dealing with statistics of the lensing convergence
in this paper, it is instructive to compare the effective
resolution of the lensing convergence with that of the lensing
excursion angle.
The lensing convergence is computed by solving the evolution equation
of the Jacobian matrix, eq.~(\ref{jacobi}).
The distance combination appearing in this equation is
$f(\chi_i)f(\chi_n-\chi_i)/f(\chi_n)$, and is shown in the middle
panel in Figure \ref{fig:resolution}.
This distance combination has a peak at intermediate redshift
depending on the source redshift.
Therefore, the worse resolution at low redshift lens planes has only
small impact on the effective resolution of lensing
signals, $\kappa$ and $\gamma$, except for cases of very low source redshifts
($z_s\la0.5$).  
Figure \ref{fig:tc} shows that the two-point correlation function of
the lensing convergence defined by $C_\kappa (\theta)=\langle \kappa(\phi)
\kappa(\phi+\theta)\rangle$ compared with the semi-analytic prediction
(Bartelmann \& Schneider 2001).
The measurements agree well with the predictions down to about 2
arcmin.
Therefore, the effective resolution of the lensing convergence is
found to be much better than that of the lensing excursion angle.

\section{Statistical properties of lensing lensing excursion angles}
\label{sec:deflection}

\begin{figure}
\begin{center}
\begin{minipage}{8.5cm}
\begin{center} 
\epsfxsize=8.5cm \epsffile{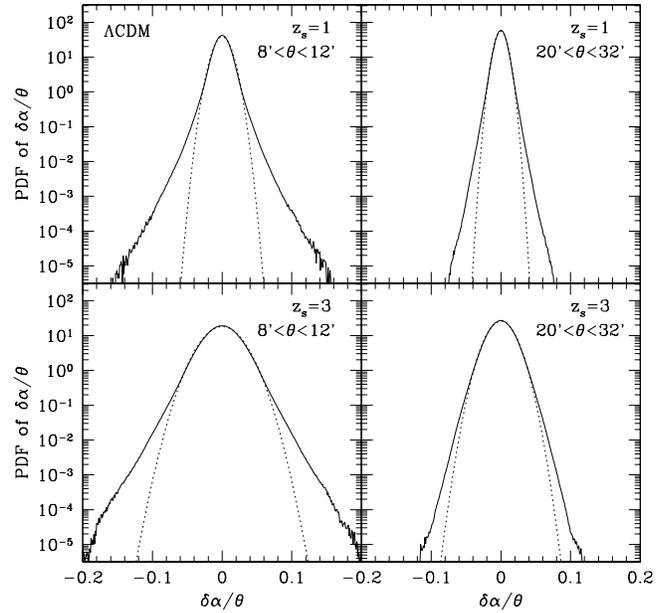}
\end{center}
\end{minipage}
\caption{The probability distribution function of the lensing excursion angle
normalized by its intrinsic separation (the solid curves).
Top panels are for $z_s=1$ and bottom panels for $z_s=3$. 
Left and right panels are for light ray pairs with intrinsic
separation $8\arcmin<\theta<12\arcmin$ and
$20\arcmin<\theta<32\arcmin$, respectively.
The dotted curves show Gaussian distribution with the $\sigma$
computed the measured PDFs}.
\label{fig:pdf}
\end{center}
\end{figure}

Figure \ref{fig:pdf} shows the probability distribution function (PDF) of
the lensing excursion angles normalized by its intrinsic separation.
Since the vector field $\bmath{\delta \alpha}(\bmath{\phi})$ has no special
direction, we take both two components, $\delta \alpha_1$ and $\delta
\alpha_2$, to compute the PDFs.
It is clearly shown in Figure \ref{fig:pdf} that the PDFs consist of
two contributions: A Gaussian distribution at inner part, and the
exponential tail at the outer part.
The inner part of the PDFs are fitted reasonably well by a Gaussian
distribution with the $\sigma$ computed from the PDFs
(root-mean-square computed from the PDFs was used for $\sigma$).
We should here noticed that the measured standard deviation agrees
very well with the semi-analytical prediction as shown in Figure
\ref{fig:deltheta}. 

The fact that the PDFs consist of two contributions suggests that
there are two different processes that make the lensing excursion angle.
One, which makes Gaussian distribution, might be secular small
(random) deflections due to linear density fluctuations along each
light ray path.
This is explained by the fact that the lensing deflection angles,
which are due to Gaussian random fluctuations (such as the linearly
evolved density fluctuation field), are Gaussian random field.
As the light ray travels longer distance, the ray can undergo
more fluctuations.
Therefore the width of Gaussian distribution increases as the source
redshift becomes higher.
The other, which makes the exponential tail, might be a single (or
possibly multiple) coherent scatter by a nonlinear structure such as
a galaxy or a cluster of galaxies\footnote{Computations of a lensing
optical depth of strong lensing events by a galaxy or a clusters of
galaxies as well as a small probability of finding multiple imaged
QSOs in a QSO catalog tell us that light rays which undergo
strong lensing more than twice times are very rare (e.g., Chiba \&
Futamase 1999; Hattori, Kneib \& Makino 1999, Hamana, Martel \&
Futamase 2000).}.
As the separation of a light ray pair decreases, smaller scale, strongly
nonlinear structures can contribute to a coherent scatter.
Therefore the exponential tail becomes more prominent for a small
separation case than a larger one as shown in Figure \ref{fig:pdf}.

Does the exponential tail make a significant influence on studies of
weak lensing effects on the CMB ?
As far as the weak lensing effects on the CMB power spectrum
concerned, it has no effect, because the crucial assumption that the
lensing excursion angle is (in a statistical sense) much smaller than the
intrinsic separation is true even in the presence of the exponential
tail. 
This can not be directly demonstrated by our simulations as
the light rays are not followed up to LSS, but can be proved by the
semi-analytic prediction which tells that the root-mean-square of the
lensing excursion angle is much smaller than the intrinsic separation angle.
The validity of the semi-analytic prediction was supported by our
numerical simulation (Figure \ref{fig:deltheta}).

\begin{figure}
\begin{center}
\begin{minipage}{8cm}
\begin{center} 
\epsfxsize=8cm \epsffile{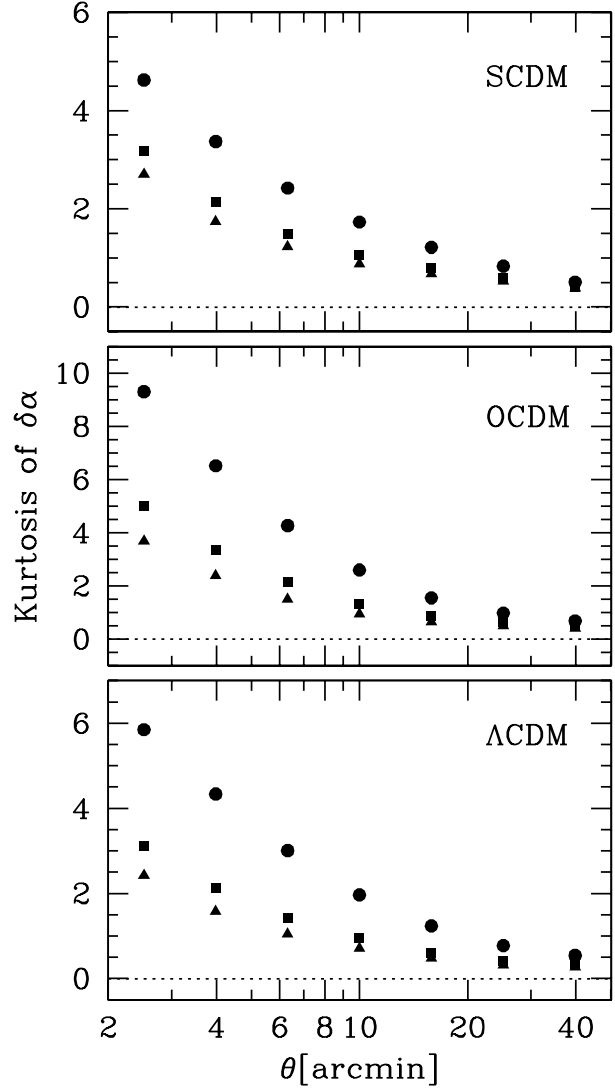}
\end{center}
\end{minipage}
\caption{Kurtosis of lensing excursion angle $\delta \alpha$, that is,
$\langle \delta \alpha^4 \rangle/\langle \delta \alpha^2 \rangle^2
-3$, as a function of the intrinsic separation of light ray pairs.
Filled circles, filled squares and filled triangles are for $z_s=1$,
$z_s=2$ and $z_s=3$, respectively.}
\label{fig:kurt_da}
\end{center}
\end{figure}

Figure \ref{fig:kurt_da} shows the Kurtosis of the PDF of $\delta
\alpha$ defined by $\langle \delta \alpha^4 \rangle/\langle \delta
\alpha^2 \rangle^2-3$.
The Kurtosis decreases as the source redshift becomes higher.
On scales larger than 10 arcmin (below that, the lensing effects on
the CMB will be hardly detected) the Kurtosis less than unity at $z_s=3$,
and thus, at the LSS, it must be smaller than that.
It can be, therefore, said that the exponential tail or, more
generally, the non-Gaussian nature in the PDF is very unlikely to have
a significant effect on the weak lensing of the CMB.

\section{Summary}
\label{sec:conclusion}

We performed ray-tracing simulation combined with sets of large
$N$-body simulations.
The use of the tiling technique enable us to explore a wide angular
dynamic range with the efficient PM $N$-body code.
The angular resolution is limited by the spatial resolution that
$N$-body simulations have.
Since the minimum box size is constrained to keep large scale modes that
may contribute to the weak lensing, the angular resolution of the low
redshift boxes becomes worse.
We found that the tiling technique particularly suits for study
of the cosmic shear statistics, because the worse angular resolution
does not make a serious impact on an effective angular
resolution of the lensing convergence and shear.
On the other hand, this worse resolution make a stronger effect on
the effective resolution of the lensing excursion angles.
One possible solution to improve the resolution is to use a higher
resolution simulation such like P$^3$M code for a few lowest redshift
boxes.

We have numerically examined statistical properties of the lensing excursion
angles.
We found that the variance in the lensing excursion angles predicted
by the power spectrum approach is in good agreement with our numerical
results.
We found that the PDF of the lensing excursion angles is not simply
Gaussian but has an exponential tail.
However, it can be safely concluded that the exponential tail, or more
generally non-Gaussian nature in the PDF has no significant effects on
the weak lensing of the CMB. 
These results demonstrate a validity of using the power spectrum
approach to compute lensing effects on the CMB.
Furthermore our results support the validity of a simple numerical
simulation method to obtain a lensed CMB map in which both the 
lensing displace filed and the CMB temperature map are generated
assuming the Gaussian statistics and the lensed CMB map is obtained by a
mapping (e.g., Takada \& Futamase 2001).

\section*{Acknowledgments}
We acknowledge that the $N$-body simulations in this work were
performed by S.~Colombi, we thank him for his collaboration and for
useful comments.
We would like to thank M.~Takada for fruitful discussions and an
anonymous referee for a detailed report that have contributed to
improvement of the manuscript.
This research was supported in part by the Direction de la Recherche
du Minist{\`e}re Fran{\c c}ais de la Recherche. 
The computational means (CRAY-98) to do the $N$-body simulations were
made available to us thanks to the scientific council of  the Institut
du D\'eveloppement et des Ressources en Informatique  Scientifique
(IDRIS).
Numerical computation in this work was partly carried out at
IAP at the TERAPIX data center and on MAGIQUE (SGI-02K).


\bsp 
\label{lastpage}

\end{document}